\documentclass[a4paper,12pt]{article}
\usepackage{float,amsthm,amsmath,amssymb}
\usepackage{url,color}

\usepackage[dvipdfmx]{graphicx}
\usepackage{tikz}
\usetikzlibrary{positioning, intersections, calc, arrows.meta,math}

\theoremstyle{plain}
\newtheorem{thm}{Theorem}
\newtheorem{prop}[thm]{Proposition}
\newtheorem{lem}[thm]{Lemma}

\newtheorem{prob}[thm]{Problem}
\newtheorem{clm}[thm]{Claim}
\theoremstyle{definition}
\newtheorem{defn}[thm]{Definition}

\newtheorem{rem}[thm]{Remark}

\begin{document}

\title{Large induced subgraph with a given pathwidth in outerplanar graphs
}
\author{Naoki Matsumoto\thanks{University of the Ryukyus, Okinawa, Japan, 
E-mail: {\tt naoki.matsumo10@gmail.com}},
Takamasa Yashima\thanks{Kanazawa Institute of Technology, Ishikawa, Japan,
E-mail: {\tt takamasa.yashima@gmail.com}}
and
Hikaru Yokoi\thanks{Keio University, Kanagawa, Japan,
E-mail: {\tt pandorabox0720@keio.jp}}
}
\date{}
\maketitle

\begin{abstract}
A long-standing conjecture by Albertson and Berman in 1979 states that 
every planar graph of order $n$ has an induced forest with at least $\lceil \frac{n}{2} \rceil$ vertices.
As a variant of this conjecture,
Chappell conjectured that 
every planar graph of order $n$ has an induced linear forest with at least $\lceil \frac{4n}{9} \rceil$ vertices.
As a partial solution to the conjecture, 
Pelsmajer in 2004 proved that every outerplanar graph of order $n$ has 
an induced linear forest with at least $\lceil \frac{4n+2}{7}\rceil$ vertices and this bound is sharp.
In this paper, 
we investigate the order of induced subgraphs with a given pathwidth in outerplanar graphs.
The above result of Pelsmajer implies that every outerplanar graph of order $n$ has 
an induced subgraph with pathwidth at most 1 and at least $\lceil \frac{4n+2}{7}\rceil$ vertices.
We extend this to obtain a result on the maximum order of induced subgraphs with a given pathwidth in an outerplanar graph.
We also give its upper bound, which generalizes Pelsmajer's construction.
\end{abstract}



\section{Introduction}

In this paper, we consider only finite, simple and undirected graphs unless otherwise stated. 
For terminology not defined in this section, see Section~\ref{sec:def}.

Albertson and Berman~\cite{AB1979} conjectured that
every planar graph has an induced forest with at least half of its vertices,
where the lower bound is best possible if it is true~\cite{AW1987}.
It is worth noting that, if this conjecture holds, 
then we can give a proof of the fact that every planar graph of order $n$ contains an independent set with at least $\lceil\frac{n}{4}\rceil$ vertices without using the Four Color Theorem.
The best known lower bound is $\frac{2n}{5}$,
which is a consequence of the acyclic 5-colorability of planar graphs by Borodin~\cite{Borodin1979}.
On the other hand,
the above conjecture is known to hold for several subclasses of planar graphs;
for example, outerplanar graphs~\cite{H1990} and triangle-free planar graphs~\cite{DMP2019}.

A similar problem has been well-studied for an induced {\it linear forest}, 
which is a forest consisting only of paths.
Inspired by Albertson and Berman's Conjecture,
Chappell conjectured that every planar graph of order $n$ has an induced linear forest with at least $\lceil \frac{4n}{9} \rceil$ vertices (cf.~\cite{P2004}).
Towards the solution to Chappell's Conjecture,
Pelsmajer~\cite{P2004} proved that 
every outerplanar graph of order $n$ has an induced linear forest with at least $\lceil \frac{4n+2}{7}\rceil$ vertices,
and this bound is best possible.
Chappell and Pelsmajer investigated this problem for graphs with a given {\it treewidth}
and proved the following theorem.

\begin{thm}[\cite{CP2013}]\label{thm:01}
Let $k,d \ge 2$ be integers.
If $G$ is a graph of order $n$ with treewidth at most $k$, then $G$ has an induced forest with maximum degree at most $d$ and at least
$\left\lceil \frac{2dn+2}{kd+d+1} \right\rceil$ vertices,
unless $G \in \{K_{1,1,3},K_{2,3}\}$ and $k=d=2$.
\end{thm}

In this paper, focusing on outerplanar graphs,
we investigate the maximum order of induced subgraphs with a given {\it pathwidth}.
To this end, we introduce a new invariant $M_{k,\ell}$ as follows: 
Let $G$ be a graph, and $k$ and $\ell$ be positive integers. 
We define $I_{\ell}(G)$ as the maximum order of induced subgraphs of $G$ with pathwidth at most $\ell$,
and $M_{k,\ell}$ as the maximum number $t$ such that 
every graph of order $t$ with treewidth at most $k$ 
has pathwidth at most $\ell$.

\begin{rem}
Note that for any positive integer $k$ and $\ell$, the value $M_{k,\ell}$ is well-defined, i.e., it is finite,
since there is a tree with pathwidth (at least) $\ell+1$.
Moreover,
it is well-known~\cite{KS1993} that 
the pathwidth of a graph of order $n$ with treewidth $k$ is $O(k \log n)$.
Thus, $M_{k,\ell} = \Omega(e^{\frac{c\ell}{k}})$ for some constant $c$.
\end{rem}

We first give the following lower bound for $I_{\ell}(\cdot)$.

\begin{prop}\label{thm:main1}
Let $k$ and $\ell$ be positive integers and let $G$ be a graph of order $n$ with treewidth at most $k$.
Then, $I_{\ell}(G) \geq \frac{M_{k,\ell}}{M_{k,\ell}+k+1}n$.
\end{prop}

This proposition follows from the following result due to Knauer and Ueckerdt~\cite{KU2023+} 
(the proof is presented in Section~\ref{sec:result}).
They focus on another conjecture by Chappell and Pelsmajer concerning treewidth and the order of induced matchings~\cite{CP2013}.
To approach the conjecture,
they consider a \emph{$c$-clustered set},
which is a vertex subset $S$ such that
every component in the subgraph induced by $S$ has at most $c$ vertices.
This notion primarily appears in the context of  \emph{clustered coloring}; see a survey~\cite{Wood2018}.

\begin{prop}[\cite{KU2023+}]\label{prop:KU}
Let $k$ and $c$ be positive integers.
If $G$ is a graph of order $n$ with treewidth at most $k$,
then there exists a $c$-clustered set 
$S$ with $|S| \ge \frac{c}{k+c+1}n$.
\end{prop}

In what follows, 
we focus only on outerplanar graphs.
We here define $M^o_{\ell}$ to be the maximum number $t$ such that every \emph{outerplanar} graph of order $t$ has pathwidth at most $\ell$.
Observe that $M^o_{\ell} \ge M_{2,\ell}$
since every outerplanar graph has treewidth at most 2.
Noting that $M^o_{2}=5$, 
we can improve the lower bound in Proposition~\ref{thm:main1} as follows when $\ell=2$.

\begin{thm}\label{thm:p2}
Let $G$ be an outerplanar graph of order $n$.
Then, $I_2(G) \ge \frac{5n}{7}$.
\end{thm}

We also give an upper bound for $I_{\ell}(\cdot)$ using $M^o_{\ell}$.

\begin{thm}\label{thm:main2}
For any positive integer $\ell$,
there is an infinite family of outerplanar graphs 
$(G_i)_{i \ge 1}$ such that each $G_i$ satisfies
$\frac{I_{\ell}(G_i)}{|V(G_i)|} = \frac{M^o_{\ell}}{M^o_{\ell}+\frac{4}{3}}$.
Similarly,
for any positive integers $k$ and $\ell$,
there is an infinite family of graphs 
$(H_i)_{i \ge 1}$
such that each $H_i$ satisfies $tw(H_i) = k$ and
$\frac{I_{\ell}(H_i)}{|V(H_i)|} = \frac{M_{k,\ell}}{M_{k,\ell}+\frac{4}{3}}$.
\end{thm}

By summarizing the above results,
we have the following table for small values of $\ell$.
In the table, 
if $\ell = 1$,
then the lower bound $\frac{4}{7}$ by Pelsmajer~\cite{P2004} is
better than ours.
Note that $M^o_{1} = 2$ and $M^o_{2} = 5$; see Figure~\ref{fig:examplesfor12}.
The upper bound of $M^o_{3}$ can be obtained by applying Proposition~\ref{prop2} to $\ell=3$.

\begin{center}
\begin{tabular}{|c|c|c|c|} \hline
$\ell$ & $M^o_{\ell}$ & Lower bound (Prop~\ref{thm:main1} \& Thm~\ref{thm:p2})& 
Upper bound (Thm~\ref{thm:main2})\\ \hline
1 & 2 & 2/5 (4/7~\cite{P2004}) &  3/5 \\ \hline
2 & 5 & 5/7  &  15/19  \\ \hline
3 & $\le 11$ &  $ \le 11/14 $ &  $ 33/37 $   \\ \hline
\end{tabular}
\end{center}

\begin{figure}[htb]
\centering
\unitlength 0.1in
\begin{picture}( 28.8000, 12.8000)(  7.6000,-18.4000)
%
\special{pn 8}%
\special{sh 1.000}%
\special{ar 800 1600 40 40  0.0000000 6.2831853}%
%
\special{pn 8}%
\special{sh 1.000}%
\special{ar 1200 1000 40 40  0.0000000 6.2831853}%
%
\special{pn 8}%
\special{sh 1.000}%
\special{ar 1600 1600 40 40  0.0000000 6.2831853}%
%
\special{pn 8}%
\special{sh 1.000}%
\special{ar 3000 600 40 40  0.0000000 6.2831853}%
%
\special{pn 8}%
\special{sh 1.000}%
\special{ar 2400 1000 40 40  0.0000000 6.2831853}%
%
\special{pn 8}%
\special{sh 1.000}%
\special{ar 2400 1400 40 40  0.0000000 6.2831853}%
%
\special{pn 8}%
\special{sh 1.000}%
\special{ar 3000 1800 40 40  0.0000000 6.2831853}%
%
\special{pn 8}%
\special{sh 1.000}%
\special{ar 3600 1400 40 40  0.0000000 6.2831853}%
%
\special{pn 8}%
\special{sh 1.000}%
\special{ar 3600 1000 40 40  0.0000000 6.2831853}%
%
\special{pn 8}%
\special{pa 3600 1000}%
\special{pa 3000 600}%
\special{fp}%
%
\special{pn 8}%
\special{pa 3000 600}%
\special{pa 2400 1000}%
\special{fp}%
%
\special{pn 8}%
\special{pa 2400 1000}%
\special{pa 2400 1400}%
\special{fp}%
%
\special{pn 8}%
\special{pa 2400 1400}%
\special{pa 3000 1800}%
\special{fp}%
%
\special{pn 8}%
\special{pa 3000 1800}%
\special{pa 3600 1400}%
\special{fp}%
%
\special{pn 8}%
\special{pa 3600 1400}%
\special{pa 3600 1000}%
\special{fp}%
%
\special{pn 8}%
\special{pa 3600 1000}%
\special{pa 2400 1000}%
\special{fp}%
%
\special{pn 8}%
\special{pa 2400 1000}%
\special{pa 3000 1800}%
\special{fp}%
%
\special{pn 8}%
\special{pa 3000 1800}%
\special{pa 3600 1000}%
\special{fp}%
%
\special{pn 8}%
\special{pa 1200 1000}%
\special{pa 800 1600}%
\special{fp}%
%
\special{pn 8}%
\special{pa 800 1600}%
\special{pa 1600 1600}%
\special{fp}%
%
\special{pn 8}%
\special{pa 1600 1600}%
\special{pa 1200 1000}%
\special{fp}%
\end{picture}%
\caption{The smallest outerplanar graphs with pathwidth 2 (left) and 3 (right).}
\label{fig:examplesfor12}
\end{figure}

In the end of the introduction, we pose the following problem.

\begin{prob}
For any positive integer $\ell$, is it true that $M^o_{\ell} = M_{2,\ell}$?
\end{prob}

The rest of this paper is organized as follows.
In the next section, 
we introduce some definitions necessary to state our results
and present some propositions related to the value of $M_{\ell}^o$.
In Section~\ref{sec:result}, we prove our main results~(Proposition~\ref{thm:main1} and Theorems~\ref{thm:p2} and~\ref{thm:main2}).
In Section~\ref{sec:result2}, we prove some propositions~(Propositions~\ref{prop1} and \ref{prop2}).
In the final section, we provide some remarks on planar graphs.

\section{Preliminaries}\label{sec:def}

For a graph $G$, $V(G)$ and $E(G)$ denote the sets of vertices and edges of $G$,
respectively.

\begin{defn}[Treewidth]\label{def:tw}
A {\it tree decomposition} of a graph $G$ is 
a pair $(T, \mathcal{X})$, 
where $T$ is a tree and $\mathcal{X}=(X_t)_{t\in V(T)}$ is a family of subsets of $V(G)$ indexed by the vertices of $T$,
such that

\begin{enumerate}
\item[(i)] $\bigcup_{t \in V(T)} X_t = V(G)$,

\item[(ii)]
for each edge $uv$ of $G$, there is an index $t \in V(T)$ such that 
$\{u,v\}\subseteq X_t$, and

\item[(iii)]
for any two vertices $t,t''$ of $T$,
$X_t \cap X_{t''} \subseteq X_{t'}$ for every vertex $t' \in V(T)$ in the $tt''$-path of $T$. 
\end{enumerate}

We call such a subset $X_t$ a {\it bag} in a tree decomposition.
The {\it width} of a tree decomposition $(T, \mathcal{X})$ is defined as $\max_{t \in V(T)}|X_t| - 1$,
and the {\it treewidth} of $G$, denoted by $tw(G)$, is the minimum width of any tree decomposition of $G$.
\end{defn}

Intuitively,
treewidth is an invariant which measures how close a given graph is to a tree.
Robertson and Seymour~\cite{RS1} introduced a variant of treewidth, 
called \emph{pathwidth}, which is defined as follows:

\begin{defn}[Pathwidth]\label{def:pw}
A {\it path decomposition} of a graph $G$ is 
a pair $(P, \mathcal{X})$, 
where $P$ is a path with vertex sequence $1,2,\dots,n$
and $\mathcal{X}=(X_i)_{i\in V(P)}$ is a family of subsets of $V(G)$ indexed by the vertices of $P$,
such that

\begin{enumerate}
\item[(i)] $\bigcup_{i \in V(P)} X_i = V(G)$,

\item[(ii)]
for each edge $uv$ of $G$, there is an index $i \in V(P)$ such that 
$\{u,v\}\subseteq X_i$, and

\item[(iii)]
for any three indices $i \le j \le k$, $X_i \cap X_{k} \subseteq X_{j}$.
\end{enumerate}

The {\it width} of a path decomposition is defined as $\max_{i \in V(P)}|X_i| - 1$,
and the {\it pathwidth} of $G$, denoted by $pw(G)$, is the minimum width of any path decomposition of $G$.
\end{defn}

Similarly to treewidth,
pathwidth measures how close a graph is to a path.
In a path decomposition of a graph $G$,
an {\it end bag} is the first or last bag in the path decomposition.

Treewidth and pathwidth are fundamental graph invariants.
A typical problem is to compute these values
(or to construct a tree or path decomposition with minimum width).
However, these problems are known to be NP-hard, 
and they remain so even when restricted to specific graph classes.
For more details on treewidth and pathwidth,
see~\cite{Bodl1993},
and for algorithmic aspects of them,
see a recent paper~\cite{GJNW2023}.

For later use, we collect the relevant notation and parameters.

\begin{defn}\label{def2}
Let $G$ be a graph, and let $\ell$ be a positive integer.

\begin{itemize}
\item $I_{\ell}(G) := \max\{|V(H)| : 
\mbox{$H$ is an induced subgraph of $G$ with pathwidth at most $\ell$}\}$.

\item $M^T_{\ell} := \max\{t : \mbox{every tree of order $t$ has pathwidth at most $\ell$}\}$.

\item $M^o_{\ell} := \max\{t : \mbox{every outerplanar graph of order $t$ has pathwidth at most $\ell$}\}$.
\end{itemize}
\end{defn}

Using Scheffler's characterization~(Lemma~\ref{lem:Scheffler} in  Section~\ref{sec:result2}) of pathwidth of trees,
we can determine the the exact value of $M^T_{\ell}$.

\begin{prop}
\label{prop1}
For any positive integer $\ell$,
we have $M_{\ell}^T=\frac{1}{2}\left(5\cdot3^{\ell}-3\right)$.
\end{prop}

The value $M^T_{\ell}$ can be used to provide a lower bound for $M^o_{\ell}$. 
In fact, we prove the following proposition.

\begin{prop}
\label{prop2}
For any integer $\ell\ge 3$,
we have 
\[ M_{\lfloor\frac{\ell-1}{2}\rfloor}^T+2 \le M_{\ell}^o\le \frac{1}{2}\left(3^{\lceil\frac{\ell+3}{2}\rceil}-5\right). \]
In particular, $M_\ell^o=\Theta(3^{\frac{\ell}{2}})$.
\end{prop}

This concludes our discussion on $M_\ell^o$;
we now return to our main results on $I_{\ell}(\cdot)$.

\section{Proof of main results}\label{sec:result}

We first prove Proposition~\ref{thm:main1} and Theorem~\ref{thm:p2}.
When $G$ is a maximal outerplane graph, 
an {\it internal face} (or {\it internal triangle}) of $G$ is a face of $G$ such that
each edge of the boundary of the face does not lie on the boundary of the infinite face of $G$.

\begin{proof}[Proof of Proposition~$\ref{thm:main1}$]
Let $G$ be a graph of order $n$ with treewidth $k$.
By applying Proposition~\ref{prop:KU} for $c = M_{k,\ell}$,
we have 
$S \subseteq V(G)$ with $|S| \ge \frac{M_{k,\ell}}{k+M_{k,\ell}+1}n$ such that 
each component of $G[S]$ has at most $M_{k,\ell}$ vertices.
By the definition of $M_{k,\ell}$, 
each component of $G[S]$ has pathwidth at most $\ell$,
and hence the pathwidth of $G[S]$ is also at most $\ell$.
Thus, $I_{\ell}(G) \ge |S| \ge \frac{M_{k,\ell}}{k+M_{k,\ell}+1}n$.
\end{proof}

\begin{proof}[Proof of Theorem~$\ref{thm:p2}$]
We prove the theorem by induction on the number of vertices.
If $|V(G)| \le 7$, then the theorem follows from $M^o_2 = 5$.
Thus, we assume $|V(G)| \ge 8$.

We first prepare the following claim.

\begin{clm}\label{clm:01}
Let $D$ be an outerplanar graph of order at most $5$ and let $v \in V(D)$.
Then, $D$ has a path decomposition with width at most $2$ such that an end bag contains $v$.
\end{clm}
\begin{proof}
It suffices to prove the claim for a maximal outerplanar graph $D$ of order 5.
Such a graph is uniquely determined up to isomorphism,
which has three triangular faces $uxy,uyz$ and $uzw$.
It is easy to see that $D$ has a path decomposition with three consecutive bags $\{u,x,y\},\{u,y,z\}$ and $\{u,z,w\}$.
Thus the claim holds.
\end{proof}

We may assume that $G$ is maximal outerplanar.
We choose an edge $ab$ of $G$ so that for a component $H$ in $G-\{a,b\}$,
$|V(H)|$ is as small as possible under $|V(H)| \ge M^o_{2} = 5$.
If $|V(H)| = 5$,
then by inductive hypothesis, we have
$I_2(G) \ge I_2(G - \{a,b\} - V(H)) + I_2(H) \ge \frac{5(n-7)}{7} + 5 = \frac{5n}{7}$.
Hence, we may assume that $|V(H)| > 5$.

Now, by the choice of $H$, there is an internal face $abc$ in the subgraph induced by $V(H) \cup \{a,b\}$ with the following properties:
\begin{itemize}
\item the component $H_1 \subseteq H$ in $G - \{a,c\}$ has order $|V(H_1)| < 5$;
\item the component $H_2 \subseteq H$ in $G - \{b,c\}$ has order $|V(H_2)| < 5$. 
\end{itemize}

Let $G' = G - \{a,b,c\} - V(H)$.
By inductive hypothesis, $I_2(G') \ge \frac{5|V(G')|}{7}$.
Moreover, by Claim~\ref{clm:01}, both the subgraphs of $G$ induced by $V(H_1) \cup \{c\}$ and by $V(H_2) \cup \{c\}$ have path decompositions with width at most $2$ such that $c$ is contained in those end bags.
Thus, $pw(H)\le 2$.
Therefore, we have 
$I_2(G) \ge I_2(G') + I_2(H) = I_2(G') + |V(H)| 
\ge \frac{5(n-|V(H)|)}{7} + |V(H)| > \frac{5n}{7}$.
\end{proof}

Next, we prove Theorem~$\ref{thm:main2}$.

\begin{proof}[Proof of Theorem~$\ref{thm:main2}$]

We prove the former statement 
(the latter is similarly proved by replacing an outerplane graph $O_s$ with a graph of order $M_{k,\ell}+1$ with treewidth $k$ and pathwidth $\ell+1$).
We construct an outerplane graph $J$ satisfying the condition.
The graph $J$ is obtained from three maximal outerplane graphs $O_1,O_2$ and $O_3$ of order $M^o_{\ell}+1$ with pathwidth $\ell+1$,
and an additional vertex $x$ joined two adjacent vertices of $O_s$ for each $s \in \{1,2,3\}$,
where the two vertices lie on the boundary of the infinite face of $O_s$.
Note that $|V(J)| = 3M^o_{\ell}+4$. 
For each $i$, let $G_i$ be $i$ disjoint copies of $J$.

Let $F$ be an induced subgraph of $J$ with $pw(F)\le \ell$.
It suffices to prove that $|V(F)|= 3M^o_{\ell}$,
which implies that $I_{\ell}(G_i) = 3iM^o_{\ell}$.
Suppose to the contrary that $|V(F)| \ge 3M^o_{\ell}+1$.
By the definition of $J$ and $O_s$,
any induced subgraph $D$ in $J$ with $pw(D) \le \ell$ has at most $M^o_{\ell}$ vertices of each $O_s$,
and so, $F$ contains exactly $M^o_{\ell}$ vertices of each $O_s$ and the vertex $x$;
thus, $|V(F)| = 3M^o_{\ell}+1$. 
For each $s\in \{1,2,3\}$, we write $F\cap O_s$ as $F_s$. 

Let $(P, (X_j)_{j\in V(P)})$ be a path decomposition of $F$ with width $\ell$.
Since $pw(F_s)=\ell$ for $s\in \{1,2,3\}$, we can find a bag $X_{j_s}\subseteq V(F_s)$ of size $\ell+1$.
By symmetry, we may assume that $j_1<j_2<j_3$.
Take $u\in X_{j_1}$ and $v\in X_{j_3}$.
It is well-known that there is no path between $u$ and $v$ in $F-X_{j_2}$ (see Lemma 12.3.1 in the Diestel's book~\cite{diestel}).
However, we find a $uv$-path in $F$ through $x$ which avoids $X_{j_2}\subseteq V(F_2)$.
This is a contradiction.  
\end{proof}


\section{Proof of Propositions~\ref{prop1} and \ref{prop2}}\label{sec:result2}

In this section, we prove Propositions~\ref{prop1} and \ref{prop2}. 
For a proof of Proposition~\ref{prop1},
we use the following lemma due to Scheffler~\cite{Sche1990}.

\begin{lem}[Scheffler~\cite{Sche1990}]
\label{lem:Scheffler}
Let $p$ be a positive integer, and let $T$ be a tree.
Then,
$pw(T)\ge p+1$
if and only if
there is a vertex $v$ in $T$ such that $T-v$ has at least three components with pathwidth at least $p$.
\end{lem}

\begin{proof}[Proof of Proposition~$\ref{prop1}$]
Let $\ell$ be a positive integer. 
It is easy to see that $M_1^T=6$, and hence the assertion holds for $\ell=1$. 
For each $i\in\{1,2,3\}$,
let $T_i$ be a tree of order $M_{\ell}^T+1$ with pathwidth $\ell+1$. 
We construct a tree $T$ by joining a vertex of each $T_i$ and a new vertex.
Then $|V(T)|=3M_{\ell}^T+4$.
By Lemma~\ref{lem:Scheffler},
$T$ is a smallest tree with pathwidth $\ell+2$.
Thus, $M_{\ell+1}^T+1=3M_{\ell}^T+4$.
Therefore, we have $M_{\ell}^T=\frac{1}{2}(5\cdot3^{\ell}-3)$.
\end{proof}

We next move on to a proof of Proposition~\ref{prop2}. 
For a 2-connected outerplane graph $G$, a \emph{weak dual tree} of $G$ is a graph obtained from a dual graph $G^*$ by removing a vertex corresponding to the infinite face.

In order to prove Proposition~\ref{prop2},
we use the following,
which is an immediate consequence from a result due to Coudert et al.~\cite{CHS2007}.
Note that for a $2$-connected outerplane graph $G$, it is shown~\cite{BF2002} that $pw(G^*) = pw(T) + 1$, where $G^*$ is a dual graph and $T$ is a weak dual tree of $G$.

\begin{thm}[Coudert, Huc and Sereni~\cite{CHS2007}]
\label{lem:CHS}
Let $G$ be a $2$-connected outerplane graph and let $T$ be its weak dual tree.
Then, $pw(G) \le 2 pw(T)+1$.
\end{thm}

\begin{proof}[Proof of Proposition~$\ref{prop2}$]
We first prove the first inequality.
For $\ell \ge 3$, we show that every outerplanar graph $G$ of order $M_{\lfloor\frac{\ell-1}{2}\rfloor}^T+2$ has pathwidth at most $\ell$.
We may assume that $G$ is maximal.
Then, the weak dual tree $T$ of $G$ has $M_{\lfloor\frac{\ell-1}{2}\rfloor}^T$ vertices,
and thus $pw(T)\le\lfloor\frac{\ell-1}{2}\rfloor$.
By Theorem~\ref{lem:CHS},
we have
\begin{align*}
pw(G)
&\le 2pw(T)+1\\
&\le2\lfloor\frac{\ell-1}{2}\rfloor+1\\
&\le2\cdot\frac{\ell-1}{2}+1=\ell,
\end{align*}
as desired.

We next prove the second inequality.
For a proof,
we construct an infinite family $(H_i)_{i\ge 1}$ of outerplanar graphs as follows.
Let $H_1$ be a triangle.
For $i\ge2$, assume that $H_{i-1}$ has been defined.
We prepare a triangle $z_1z_2z_3$ and three disjoint copies $H_{i-1}^{(1)}, H_{i-1}^{(2)}$ and $H_{i-1}^{(3)}$ of $H_{i-1}$,
take an outer edge $x_jy_j$ of $H_{i-1}^{(j)}$ for each $j\in\{1,2,3\}$, and
join $z_j$ and each vertex in $\{x_j,y_{j+1}\}$ for each $j\in\{1,2,3\}$~(we take $y_4=y_1$).
We let $H_i$ be the resulting outerplanar graph.
By the construction, $H_i$ is $2$-connected and has $\frac{1}{2}\left(3^{i+1}-3\right)$ vertices.

\begin{clm}\label{clm:02}
Suppose that $pw(H_{i-1})=p$ for some $p>0$.
Then, $pw(H_i)\ge p+2$.
\end{clm}
\begin{proof}
Suppose to the contrary that $H_i$ has a path decomposition $(P, (X_k)_{k\in V(P)})$ of width at most $p+1$.
Since $pw(H_{i-1}^{(j)})=p$ for each $j\in\{1,2,3\}$,
there exist three distinct bags $X_{k_1}, X_{k_2}$ and $X_{k_3}$
such that $X_{k_j}$ contains at least $p+1$ vertices of $H_{i-1}^{(j)}$.
By symmetry, we may assume that $k_1< k_2 < k_3$.
Take $u\in\left(X_{k_1} \cap V(H_{i-1}^{(1)})\right)\setminus X_{k_2}$
and $v\in\left(X_{k_3} \cap V(H_{i-1}^{(3)})\right)\setminus X_{k_2}$.
Since $H_i$ is $2$-connected,
there are two internally disjoint $uv$-paths $Q_1$ and $Q_2$
with $V(Q_m)\cap\{z_1,z_2,z_3\}\neq\emptyset$ for $m\in\{1,2\}$.
Because there is no $uv$-path in $H_i-X_{k_2}$,
we have $|X_{k_2}\cap\{z_1,z_2,z_3\}|\ge 2$.
Thus, $|X_{k_2}|\ge|X_{k_2}\cap V(H_{i-1}^{(2)})|+|X_{k_2}\cap\{z_1,z_2,z_3\}|\ge (p+1)+2=p+3$,
a contradiction.
\end{proof}

By $pw(H_1)=2$ and Claim~\ref{clm:02},
we have $pw(H_i)\ge2i$.
Since $pw(H_{\lceil\frac{\ell+1}{2}\rceil})\ge2\cdot\lceil\frac{\ell+1}{2}\rceil\ge2\cdot\frac{\ell+1}{2}=\ell+1$,
we have
\begin{align*}
M_\ell^o
&\le |V(H_{\lceil\frac{\ell+1}{2}\rceil})|-1\\
&=\frac{1}{2}\left(3^{\lceil\frac{\ell+1}{2}\rceil+1}-3\right)-1\\
&=\frac{1}{2}\left(3^{\lceil\frac{\ell+3}{2}\rceil}-5\right),
\end{align*}
as desired.
\end{proof}

\section{Remarks on planar graphs}

It is well-known that 
the treewidth of planar graphs cannot be bounded by a constant 
(e.g., the grid graph, which is obtained by the Cartesian product of two paths),
whereas that of outerplanar graphs is bounded.
This implies that Theorem~\ref{thm:01} is not useful 
for estimating the upper bound of their pathwidth.
Therefore,
estimating the lower/upper bounds for
the order of maximum induced subgraph with a given pathwidth in a plane triangulation 
is very different from the outerplanar case.

On the other hand, 
it is clear
that for any graph $G$,
$I_{\ell}(G) \le I_{\ell}(H)$ for a spanning subgraph $H$ of $G$.
Moreover, $I_{\ell}(G) \geq I_{\ell-1}(G)$ for any $\ell \geq 2$.
Motivated by Chappell's Conjecture,
we propose the following problem.

\begin{prob}\label{pro:01}
For a positive integer $\ell$
and every planar graph $G$ of order $n$,
determine $I_{\ell}(G)$.
In particular, 
resolve whether or not there is a nonnegative 
and monotone increasing function $f(\ell)$ such that $I_{\ell}(G) \ge (\frac{4}{9} + f(\ell))n$.
\end{prob}

Note that it is not difficult to see that 
the latter part of Problem~\ref{pro:01} holds for $\ell\ge 7$ as follows.

\begin{prop}\label{prop3}
Let $\ell$ be a positive integer, and let $G$ be a planar graph of order $n$. 
Then, $I_\ell(G)\ge \frac{M^o_{\ell}n}{2(M^o_{\ell}+3)}$.
\end{prop}

By Propositions~\ref{prop1} and~\ref{prop2}, we see that $\frac{M^o_{\ell}}{2(M^o_{\ell}+3)}\ge \frac{4}{9}$ for $\ell\ge 7$. 
Thus, we are interested in whether the latter part of Problem~\ref{pro:01} holds for small $\ell$.

\begin{lem}\label{obs1}
Let $G$ be a planar graph. 
Then, the vertex set of $G$ can be partitioned into two subsets both of which induce outer planar graphs.
\end{lem}

\begin{proof}[Proof of Lemma~$\ref{obs1}$]
Assume that $G$ is connected. 
Fix a vertex $r$ of $G$. 
For each $i\ge 0$, let $L_i = \{v\in V(G) :\ $\rm{dist}$_G(v,r)=i \}$. 
Let $A := \bigcup_{i : \text{odd}} L_i$ and $B := \bigcup_{i : \text{even}} L_i$. 
Then, $V(G) = A\cup B$ is a desired partition. 
To see this, it is enough to show that each $G[L_i]$ is outerplanar.
Suppose to the contrary that $G[L_i]$ contains $K_4$ as a minor. 
Then, we can find a $K_5$ minor in $G$ by contractiong $G[L_1\cup L_2\cup \cdots \cup L_{i-1}]$ into a single vertex.
This contradicts the planarity of $G$. 
By the same argument, we can see that $G[L_i]$ contains no $K_{2,3}$ as a minor.
\end{proof}

\begin{proof}[Proof of Proposition~$\ref{prop3}$]
Let $G$ be a planar graph of order $n$. 
By Lemma~\ref{obs1}, we can take $A\subseteq V(G)$ with $|A|\ge \frac{n}{2}$ which induces an outerplanar graph.
For any positive integer $\ell$, by Proposition~\ref{thm:main1}, we see that $I_{\ell}(G[A])\ge \frac{M^o_{\ell}}{M^o_{\ell}+3}|A| \ge \frac{M^o_{\ell}}{2(M^o_{\ell}+3)}n$.
\end{proof}

\section*{Declaration of competing interest}

The authors declare that they have no known competing financial interests or personal relationships that could have appeared to influence the work reported in this paper.

\section*{Data availability}

No data was used for the research described in the article.

\section*{Acknowledgments}

The authors would like to thank Professor Shinji Sakamoto for his valuable comments.
This work was supported by the Research Institute for Mathematical Sciences, an International Joint Usage/Research Center located in Kyoto University. 
The first author was supported by JSPS Grant-in-Aid for Scientific Research (C) 22K11911.
The second author was supported by JSPS Grant-in-Aid for Scientific Research (C) 26K06900.
The third author was supported by  JST SPRING Japan Grant Number JPMJSP2123 and JST ERATO Grant Number JPMJER2301, Japan.

\end{document}